\documentclass[12pt]{article}
\usepackage{epsf}
\def\be{\begin{equation}}
\def\ee{\end{equation}}
\def\bea{\begin{eqnarray}}
\def\eea{\end{eqnarray}}

\begin{document}
\begin{titlepage}
\begin{center}
{\Large \bf William I. Fine Theoretical Physics Institute \\
University of Minnesota \\}
\end{center}
\vspace{0.2in}
\begin{flushright}
FTPI-MINN-16/21 \\
UMN-TH-3533/16 \\
July 2016 \\
\end{flushright}
\vspace{0.3in}
\begin{center}
{\Large \bf Two charmoniumlike charged axial resonances near 3885\,MeV 
\\}
\vspace{0.2in}
{\bf  M.B. Voloshin  \\ }
William I. Fine Theoretical Physics Institute, University of
Minnesota,\\ Minneapolis, MN 55455, USA \\
School of Physics and Astronomy, University of Minnesota, Minneapolis, MN 55455, USA \\ and \\
Institute of Theoretical and Experimental Physics, Moscow, 117218, Russia
\\[0.2in]

\end{center}

\vspace{0.2in}

\begin{abstract}
It is argued that the charged $Z^+_c(3885)$ resonance, treated as a `molecular' state of charmed $D$ and $D^*$ mesons, is likely to consist of two peaks unequally coupled to the $D^{*+} \bar D^0$ and  $D^+ \bar D^{*0}$ channels. The peaks should be split in mass by at least approximately 1.5\,MeV. This behavior arises from an enhancement of the effect of isospin violation in the masses of the $D$ and $D^*$ mesons due to apparent suppression of forces between the mesons depending on the spins of the heavy as well as of the light quarks. The suggested double-peak structure can be studied either by direct shape measurement in the channels with heavy mesons, or by isospin-violating transitions from $Z_c^\pm(3885)$ to the states of charmonium plus a light meson.

\end{abstract}
\end{titlepage}

The exotic  states $Z_b$~\cite{bellez} and $Z_c${\cite{besz1,besz2} in the charmonium and bottomonium sectors are new hadronic objects and attract a considerable interest due to their manifestly four-quark nature. The proximity of their masses to the threshold of a heavy meson-antimeson pair suggests that these are dominantly molecular objects, i.e. bound and/or resonant $S$-wave states of the heavy meson pairs. Namely, the bottomonium-like resonances $Z_b(10610)$ and $Z_b(10650)$ are respectively $B^* \bar B - B \bar B^*$ and $B^* \bar B^*$ molecules~\cite{bgmmv},  while the charmonium-like exotic states $Z_c(3885)$ and $Z_c(4020)$ are molecular resonances in the $D^* \bar D$ ($D \bar D^*$) and $D^* \bar D^*$ channels. The molecular interpretation of these states is supported by the observation of a high rate of their decay into the corresponding heavy meson pair: $Z_b(10610) \to B^* \bar B \, (B \bar B^*)$, $Z_b(10650) \to B^* \bar B^*$~\cite{bellebb2} as well as $Z_c(3885) \to D^* \bar D \, (D \bar D^*)$\cite{besz1dd} and $Z_c(4020) \to D^* \bar D^*$~\cite{besz2dd}. Furthermore, as naturally expected, the $Z_Q$ resonances come in isotopic triplets: for each of these resonances the initial discovery of electrically charged states has been followed by observation of the neutral one~\cite{bellez0,cleozc0,besz10,besz20}.

All the $Z_Q$ states have spin-parity quantum numbers $J^P=1^+$, and
in the limit of exact isotopic symmetry all theses states have definite $G$ parity equal to +1, as they are coupled to the channels with a $C$-odd state of heavy quarkonium and a pion, e.g. $Z_c \to \pi \, J\psi$, or $Z_b \to \pi \, \Upsilon$. It is clear however that combining vector ($V$) and pseudoscalar ($P$) heavy mesons in $S$ wave generally results in a broader set of the quantum numbers $I^G(J^P)$. Namely in addition to the two $1^+(1^+)$ $Z_Q$ states there are four $G$-odd combinations, named as $W_{QJ}$ in Ref.~\cite{mvwb}: $1^-(0^+): \, W_{Q0} \sim P \bar P$,  $1^-(1^+): \, W_{Q1} \sim V \bar P + P \bar V$, $1^-(0^+): \, W'_{Q0} \sim V \bar V$, and $1^-(2^+): \, W_{Q2} \sim V \bar V$. 

Some further insight into the internal dynamics of the four-quark states can be gained from analyzing the spin structure with respect to the spin of the heavy and light quarks for widely separated heavy meson pairs. Indeed, in the heavy flavored mesons the spin of the heavy quark (antiquark) is fully correlated with that of the light (anti)quark, and the spin-spin interaction splits the masses of the vector and pseodoscalar mesons. In a videly separated pair of a heavy meson and antimeson this correlation is preserved, which implies that such pair is in a mixed state with respect to the total spin $S_H$ of the heavy quark-antiquark pair as well as the spin of light one, $S_L$~\cite{bgmmv}. The spin-dependent interaction between the heavy quarks in QCD is suppressed by the inverse of the heavy quark mass, thus resulting in the heavy quark spin symmetry (HQSS). For the discussed molecular states this implies, in particular, that the interaction between the mesons does not depend (approximately) on $S_H$, but generally does depend on the spin of the light pair $S_L$. Based on this behavior and the existence of the $Z_Q$ resonances, one can generally expect~\cite{mvwb} that some or all of the $W_{QJ}$ meson pair combinations also do form molecular resonances.

It can be readily noticed that near the $V \bar P$ threshold there are in fact two states with all quantum numbers being identical, except for the $G$ parity: the lower $Z_Q$ state and $W_{Q1}$. The purpose of the present paper is to consider the effects of the mixing between the electrically charged components of these two states, arising from the $G$ parity breaking due to the isotopic mass differences of the $D$ and $D^*$ mesons. Only the charged states are being discussed here because the analogous effect in the neutral components is entirely different. Indeed the neutral antisymmetric state $Z_Q \sim V \bar P - P \bar V$ is protected from mixing with the symmetric one $W_{Q1} \sim V \bar P + P \bar V$ by the $C$ parity, and the mass difference between the charged and neutral heavy mesons results rather in a mixing of isovector states of the same charge symmetry with the isoscalar states. The isoscalar states however are affected by their mixing with pure $Q \bar Q$ quarkonium, so that there arises a more complicated dynamics of three types of states: the isovector molecule, the isoscalar molecule, and the quarkonium --- a behavior known~\cite{mvc} from the properties of the $X(3872)$ resonance. The considered here mixing between the charged components of two different isotopic triplets is significantly simpler, since it contains no effects of mixing with pure quarkonium.

In all likelihood the discussed mixing can be of an observable significance only in the charm sector, since for the $B$ and $B^*$ mesons the isotopic mass differences are very small, and in fact are not measured as of yet. For the charmed mesons the mixing is driven by the difference in the masses of the pairs $D^+ \bar D^{*0}$ and $D^{*+} \bar D^0$:~
$\mu = M(D^+)-M(D^0) - M(D^{*+}) + M(D^{*0}) = 1.46 \pm 0.11\,$MeV.  Clearly, this mass difference corresponds to a $G$-odd term in the Hamiltonian, and generally results in a mixing between states of opposite $G$ parity. In order to estimate the effects of the mixing the value of $\mu$ should be compared with the mass parameters of the  states of heavy meson pairs  and  the difference in these parameters for the states with opposite $G$ parity. 

In fact it can be argued that in the limit of isotopic symmetry there likely is a resonance $W_{c1}$ that is approximately degenerate in mass with the $Z_c$ and with a very similar width, so that any differences in the mass and width are not much larger than $\mu$.  Indeed, the splitting between the $Z_c$ and $W_{c1}$ channels is caused by the interaction between the mesons that mixes the pairs $D^+ \bar D^{*0}$ and $D^{*+} \bar D^0$: $D^+ \bar D^{*0} \to D^{*+} \bar D^0$. In the HQSS limit this interaction is the one that depends on the spin $S_L$ of the light quark pair. It has been recently argued~\cite{mvls}  that the observed~\cite{bellebb2} apparent suppression of the decay $Z_b(10650) \to B^* \bar B + c.c.$ implies a strong suppression of such interaction at shorter distances corresponding to the momentum transfer $q \approx 0.5\,$GeV. It can be also argued that, especially in the case of the charmed mesons, the spin-dependent interaction is also very weak at long distances. Indeed, the interaction at long distances is described by the one pion exchange, and in the case of the charged states it is only the exchange of the neutral pion $\pi^0$. The vertex for the $D^*D \pi^0$ interaction can be written, for nonrelativistic charmed mesons, in terms of the effective Hamiltonian
\be
H_{D^*D\pi} = {g \over f_\pi} \, \left [ {(D^+)^\dagger D^{*+}_i - (D^0)^\dagger D^{*0}_i} \right ] \, \partial_i \pi^0 + {\rm h.c.}~,
\label{ddpi}
\ee
where $f_\pi \approx 132\,$MeV is the charged pion decay constant, and $g$ is a dimensionless constant, whose value can be readily determined from the known~\cite{babar,pdg} rate of the decay $D^{*+} \to D \, \pi$: $g^2 \approx 0.15$. The exchange of the pion then results in the mixing potential, described in the $S$ wave by the expression in the momentum space
\be
\langle D^* \bar D | V(p) | D \bar D^* \rangle = {g^2 \over 3 \, f_\pi^2} \, {p^2 \over p^2 + \delta^2}~
\label{vp}
\ee
with $\delta^2 = m_\pi^2 - [M(D^*)-M(D)]^2 \approx = - (41\,{\rm MeV})^2$. In the coordinate space the long distance part of the interaction can thus be evaluated as
\be
\langle D^* \bar D | V(r) | D \bar D^* \rangle = -{g^2  \over 12 \, \pi } \, { \delta^2 \over f_\pi^2} \, {e^{-\delta \, r} \over r}
\label{vr}
\ee
which estimate for the interaction can be also read from the results of the general analysis of the pion exchange potential in Ref.~\cite{nv}. Since the parameter $\delta^2$ is negative, the potential is oscillating due to the kinematically possible decay $D^{*+} \to D^+ \, \pi^0$. Whatever the implications of the oscillations may be, numerically the coefficient in the potential is very small:
$$-{g^2  \over 12 \, \pi } \, { \delta^2 \over f_\pi^2} \approx 4 \times 10^{-4}~,$$
and the effect of this interaction on the splitting between $Z_c$ and $W_{c1}$ appears to be totally negligible in the scale of $\mu$. 

It thus can be concluded that the interaction between the heavy mesons likely does not split, in the limit of exact isotopic symmetry, the $Z_c$ and $W_{c1}$ states, and the existence of the resonance $Z_c$ implies that a  $G$-odd state $W_{c1}$ should also exist. A difference in masses and widths of these resonances can still arise from their coupling to the channels with charmonium and light mesons. Indeed the $Z_c$ decays into $\pi \, J/\psi$ and, potentially, into other similar states, e.g $\pi \, h_c$, while analogous decay channels allowed for $W_{c1}$ are $\rho \, J/\psi$, $\pi \, \chi_{cJ}$, etc. Unless there is an accidental degeneracy between these couplings (see e.g. in \cite{mvls,lv}), some difference in the widths and the masses of the resonances may arise. However this difference is likely to be numerically of the same order as $\mu$. Indeed, the measurements of the total width of $Z_c(3885)$ currently produce~\cite{pdg} the average $\Gamma_Z = 28.1 \pm 2.6\,$MeV with the decay into $D^* \bar D$ + c.c. being dominant. Thus the partial decay width into charmonium and a pion amounts to only a few MeV, and it can be reasonably expected that with similar decay rates of the $W_{c1}$ resonance, the difference in the rates is still smaller.

\begin{figure}[ht]
\begin{center}
 \leavevmode
    \epsfxsize=10cm
    \epsfbox{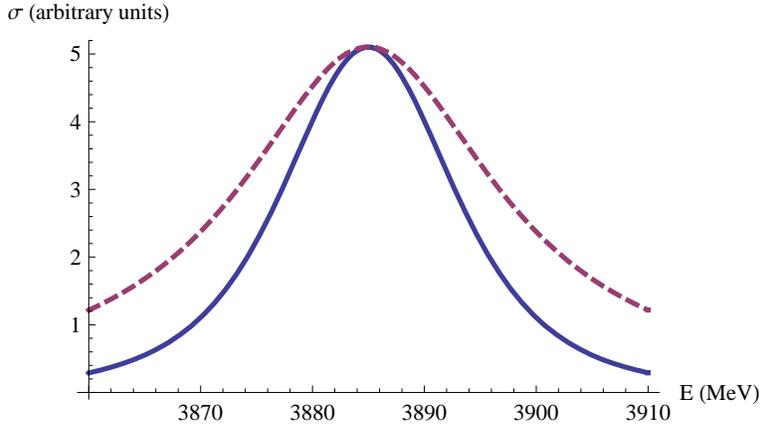}
    \caption{The shape of the dependence on the invariant mass $E$ of charmonium plus pion of the yield in the $G$ parity violating channel (solid) compared with the standard Breit-Wigner curve for the allowed channel (dashed) in the limit of degenerate $Z_c$ and $W_{c1}$ resonances. The curve for the forbidden channel is scaled by the inverse of the overall suppression factor (see text). }
\end{center}
\end{figure}

Clearly, in the limit where the difference in the masses and widths of the isotopic eigenstates $Z_c$ and $W_{c1}$ is small as compared to $\mu$, a maximal mixing takes place resulting in the true independent eigenstates of the Hamiltonian being $D^+ \bar D^{*0}$ and $D^{*+} \bar D^0$ with a resonance in each of these channel. The resonances in the two channels are split in mass by $\mu$, which splitting can be tested in the processes $e^+e^- \to \pi^- \, D^+ \bar D^{*0}$ and  $e^+e^- \to \pi^- \, D^{*+} \bar D^0$, provided that a sufficient accuracy in the measurement of the invariant mass of the heavy meson pair becomes available. Since the splitting $\mu$ between the peaks is small in comparison with their width $\Gamma$, in the processes with charmonium and a light meson in the final state that are allowed by the $G$ parity, such as the observed one $e^+e^- \to \pi \, Z_c \to \pi \, \pi \, J/\psi$, the two peaks add up and essentially fully overlap and appear as a single resonance. However 
in the processes with charmonium and a pion in the final state that are forbidden by $G$ parity, e.g. $e^+e^- \to \pi^\mp \, \pi^\pm \, \chi_{cJ}$ the Breit-Wigner factors for the two resonances come with opposite sign, and the yield should display a small but narrow peak (as illustrated in Fig.~1):
\be
\sigma (e^+e^- \to e^+e^- \to \pi^\mp \, \pi^\pm \, \chi_{cJ}) \propto {\mu^2 \over [(E-M)^2 + \Gamma^2/4]^2} ~,
\label{sgf}
\ee
where $E$ is the invariant mass of the charmonium $\chi_{cJ}$ and a pion, and $M \approx 3885\,$MeV is the mass of the discussed resonances. The parameter for the suppression of the rate of a $G$-violating process relative to a $G$-allowed one is $\mu^2/\Gamma^2 \approx 3 \times 10^{-3}$.

\begin{figure}[ht]
\begin{center}
 \leavevmode
    \epsfxsize=16cm
    \epsfbox{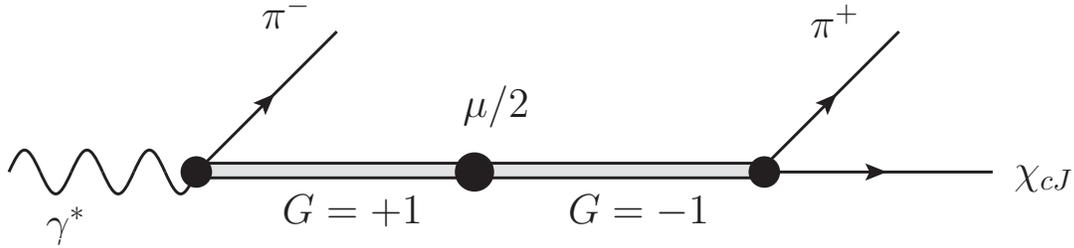}
    \caption{The graph for the $G$ parity violating process $e^+e^- \to \pi^+ \pi^- \chi_{cJ}$ in the first order in the $G$ violating mass difference $\mu$.  }
\end{center}
\end{figure}

In the opposite limit, where the distance between the pole positions of the $Z_c$ and $W_{c1}$ resonances, $M_Z - i \Gamma_Z/2$ and $M_W-i \Gamma_W/2$, can be considered as large in comparison with $\mu$, the mixing can be treated perturbatively. In this case the amplitude for $G$ parity violating processes can be evaluated from a graph shown in Fig.~2, e.g. 
\be
\sigma (e^+e^- \to e^+e^- \to \pi^\mp \, \pi^\pm \, \chi_{cJ}) \propto { \mu^2 /4 \over [ (E-M_Z)^2 + \Gamma_Z^2/4] \, [ (E-M_W)^2 + \Gamma_W^2/4]}
\label{sgf1}
\ee
with the overall scale of the suppression being of the same order as in the previous case of degenerate states, provided that the difference in masses and widths is not larger than the width of the $Z_c(3885)$. In this case the effect can be more noticeable in the energy dependence of the relative yield in the processes $e^+e^- \to \pi^- \, D^+ \bar D^{*0}$ and  $e^+e^- \to \pi^- \, D^{*+} \bar D^0$. Indeed, the amplitudes for these processes contain an interference between the $G=-1$ and $G=+1$ production amplitudes:
\bea
&&A(e^+e^- \to \pi^- \, D^+ \bar D^{*0}) \propto {1 \over E-M_Z +i \Gamma_Z/2 } \left [ 1+ {\mu/2 \over E-M_W+i \Gamma_W/2} \right ], \nonumber \\
&&A(e^+e^- \to \pi^- \, D^{*+} \bar D^0) \propto - {1 \over E-M_Z +i \Gamma_Z/2 } \left [ 1- {\mu/2 \over E-M_W+i \Gamma_W/2} \right ]~,
\label{add}
\eea 
so that the asymmetry between the two final channels is linear in $\mu$:
\be
{\sigma(e^+e^- \to \pi^- \, D^+ \bar D^{*0}) - \sigma (e^+e^- \to \pi^- \, D^{*+} \bar D^0) \over \sigma(e^+e^- \to \pi^- \, D^+ \bar D^{*0}) + \sigma (e^+e^- \to \pi^- \, D^{*+} \bar D^0)} = \mu  \, {E- M_W  \over (E-M_W)^2 + \Gamma_W^2/4}~,
\label{asym}
\ee
and should display a variation across the $G$-odd $W_{c1}$ resonance. Clearly, this asymmetry is maximal at $|E-M_W| = \Gamma_W/2$, where its absolute value is $\mu/\Gamma_W$ and can amount to $\sim 5\%$, provided that $\Gamma_W$ is not much larger than $\Gamma_Z \approx 28\,$MeV.

In summary. The apparent weakness of the interaction that converts between the $J^P = 1^+$ $D^+ \bar D^{*0}$ and  $D^{*+} \bar D^0$ charmed meson pairs implies that there should be a significant degeneracy between the states of these pairs with opposite $G$ parity. In particular, the observed resonance structure $Z_c(3885)$ that is $G$-even n the limit of exact isotopic symmetry should have a $G$-odd counterpart $W_{c1}$ with very similar mass and width parameters. This approximate degeneracy enhances the effect of the isospin (and thus $G$ parity) breaking by the mass difference $\mu$ between the two meson pairs and should result in a  double resonance structure of states with mixed $G$ parity, in particular with an unequal content of the $D^+ \bar D^{*0}$ and  $D^{*+} \bar D^0$ pairs. The effects of such structure can possibly be observed in future measurements either by the asymmetry of the yield of these pairs [Eq.(\ref{asym})], or by an observation of isospin ($G$ parity) breaking transitions to charmonium plus a light meson [Eq.(\ref{sgf})]. While it can be troublesome to perform such measurements within the present experimental setting, they may become feasible in future studies.

This work is supported in part by U.S. Department of Energy Grant No.\ DE-SC0011842.

\end{document}